\documentclass[aps,prl,twocolumn,superscriptaddress,showpacs] {revtex4}

\usepackage{epsfig}
\usepackage{amssymb}

\begin{document}

\title{Steady-state conduction in self-similar billiards}
\author{Felipe Barra}
\affiliation{Deptartamento de F\'{\i}sica, Facultad de Ciencias
  F\'{\i}sicas y Matem\'aticas, Universidad de Chile, Casilla 487-3,
  Santiago Chile} 
\author{Thomas Gilbert}
\affiliation{Center for Nonlinear Phenomena and Complex Systems,
  Universit\'e Libre  de Bruxelles, CP~231, Campus Plaine, B-1050
  Brussels, Belgium}
\date{\today}
\begin{abstract}
The self-similar Lorentz billiard channel is a spatially extended
deterministic dynamical system which consists of an infinite
one-dimensional sequence of cells whose sizes increase monotonously
according to their indices. This special geometry induces a non-equilibrium
stationary state with particles flowing steadily from the small to the
large scales. The corresponding invariant measure has fractal properties
reflected by the phase-space contraction rate of the
dynamics restricted to a single cell with appropriate boundary conditions. In
the near-equilibrium limit, we find numerical agreement
between this quantity and the entropy production rate as specified by
thermodynamics. 
\end{abstract}
\maketitle


Over the last two decades, the non-equilibrium statistical mechanics of
chaotic dynamical systems has received a great deal of attention, in
particular regarding the positivity of entropy production and its
connection to dynamical properties characteristic of chaos, such as the
Lyapunov exponents \cite{H86,EM90,Gas98,Dor99}.

Of particular interest is the two-dimensional periodic Lorentz gas with
external forcing and Gaussian iso-kinetic thermostat (GIKLG)
\cite{H86}. This is an elementary model of electronic conduction under a
non-equilibrium constraint. The thermostat is actually a mechanical
constraint and acts so as to remove the energy input from the external
forcing \cite{EM90}. Under the action of this thermostat, the 
kinetic energy remains constant and thus fixes the temperature of the
system; no interaction with a hypothetical environment is needed in order
to achieve thermalization. Rather, dissipation occurs in the bulk. As
shown in \cite{CELS93}, this model enjoys strong chaotic properties. Its 
natural invariant measure is fractal, with one positive and 
one negative Lyapunov exponents, whose sum is negative, and identified as
minus the entropy production rate \cite{R96}. The comparison with 
the corresponding phenomenological expression provides a relation between
the phase space contraction rate and conductivity. 

The GIK dynamics can actually be given a Hamiltonian formulation
\cite{DM98}. This result is contained in a more general formalism
\cite{W00}, by which one identifies the GIK trajectories with the
geodesics of a torsion free connection, called the Weyl connection. In this
formalism, one shows the GIKLG is equivalent, by a conformal
transformation, to a distorted billiard whose trajectories are straight
lines, referred to as W-flow. In this geometry, the cells are stretched and
trajectories accelerated in the direction of the external field. A detailed
analysis of the W-flow \cite{BG} reveals that the cells' stretching is the
essential mechanism to driving the non-equilibrium stationary state.

In this letter, we investigate the connection between the chaotic dynamics
and non-equilibrium thermodynamics of a self-similar billiard chain, such
as defined in \cite{BG06}. This billiard chain can be thought of as a
simplified version of W-flow described above; we retain the sretched
geometry of the cells, but get rid of the bulk dissipation.

\begin{figure}
\includegraphics[width=.45\textwidth]{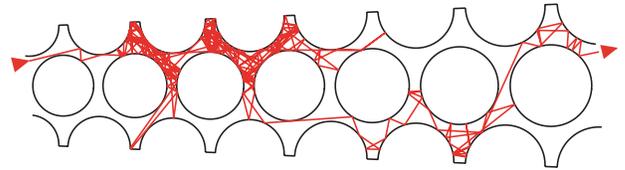}
\caption{(Color online) The self-similar Lorentz channel billiard and a
  trajectory.} 
\label{fig.channel}
\end{figure}

A self-similar billiard is thus a one-dimensional 
spatially extended system such as shown in Fig.~\ref{fig.channel}. It
consists of an infinite collection of two-dimensional cells, identical in
shapes, but differing in sizes, which are scaled and glued together along a
horizontal axis. Each cell consists of convex obstacles upon which point
particles collide elastically, independently of one another. The left and
right borders of every cell are open and scaled by a constant ratio which
we denote by $\mu$. The system is so constructed that $\mu=1$ corresponds
to the Lorentz channel \cite{Gas98}, which exhibits diffusion, but no
bias. When $\mu\ne1$, the system sustains a non-equilibrium stationary
state, characterized by a steady density current from the smaller to the
larger scales. 

Strictly speaking, this self-similar billiard channel is different from the
W-flow, constructed along the lines of \cite{W00}. However the two systems
have in common the scaling property 
between the left and right borders of every cell. An important difference
concerns the dynamics between collisions. Here we assume free propagation
of point particles between collision events, which amounts to suppressing
the thermostating mechanism of the GIK dynamics. 
Another example of a billiard with a similar scaling mechanism was
considered in \cite{BR01}. Other systems which bear similarities with
this one are the multibaker maps with energy \cite{TG99}. Both systems are 
area-preserving and energy conserving, with an external driving force (here
in the form of a geometric constraint) that induces a macroscopic current,
without the necessity of a thermostating mechanism.

In order to specify the stationary properties of the system, we resort to
the special flow of the billiard \cite{CFS82}, thus substituting 
the real time dynamics by a discrete mapping from one collision event to
the next. This is a usual procedure for the study of ergodic properties of
billiards, which allows us to replace the volume-preserving dynamics on
the extended system  by a phase-space contracting mapping on a
periodic cell. 
This feature is also common to multibaker maps with energy, which
can be reduced to dissipative baker maps such as studied in \cite{VTB97},
see \cite{TG99}. Moreover the invariant 
measure has fractal properties much like that of the GIKLG \cite{MDR97}. 

Following \cite{R96}, we identify the phase space contraction rate with the
entropy production rate, and take the limit $\mu\to1$ so as to compare this
expression to its thermodynamic counterpart. We thus infer a relation
between the current of the near-equilibrium system ($\mu\ne1$) and the
diffusion coefficient of the equilibrium system ($\mu=1$), much like a
fluctuation-dissipation theorem. This result is confirmed by our numerical
computations.

\begin{figure}[htb]
\centering{
\includegraphics[height=5cm]{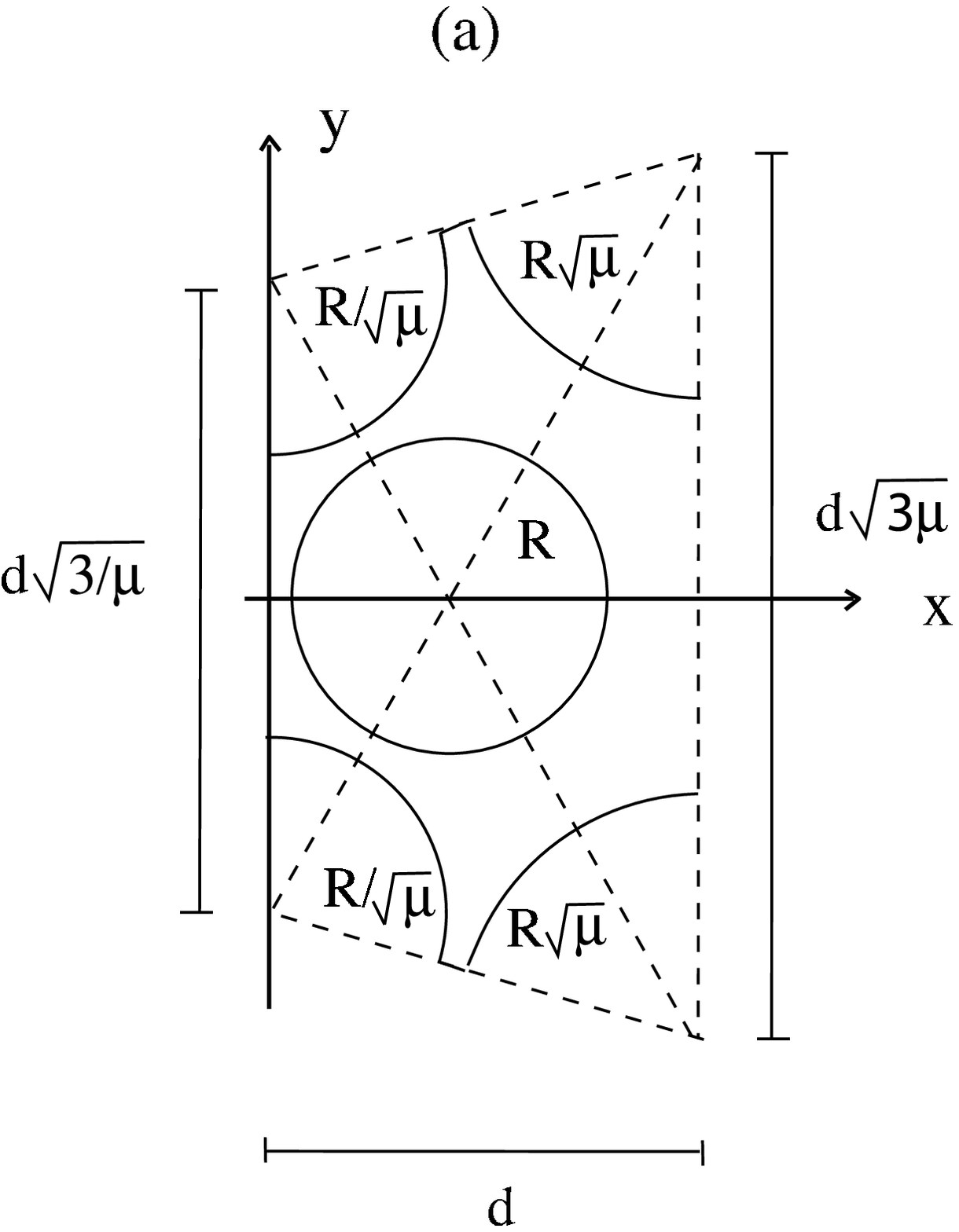}
\hspace{.5cm}
\includegraphics[height=5cm]{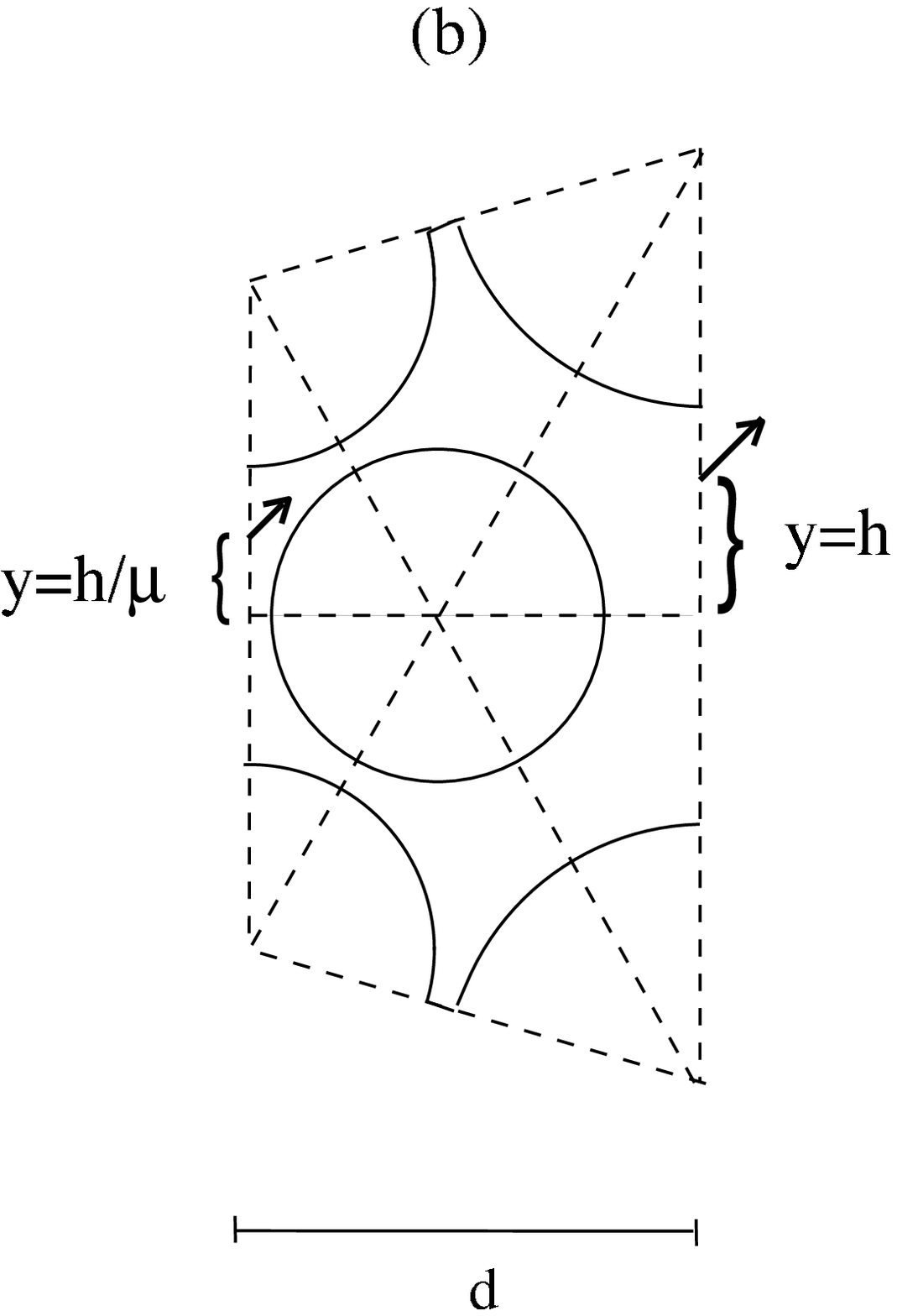}
}
\caption{(a) Geometry of the reference cell. (b) Sketch of the
  boundary conditions in the periodic cell, as given by
  Eqs.~(\ref{pbright}-\ref{pbleft}).}
\label{fig.unitcell}
\end{figure}

Going back to the definition of the model, the reference cell, with index
$i=0$, is represented in Fig.~\ref{fig.unitcell}(a). 
It is a region defined by the exterior of five disks, four of which are
centered at the corners of the cell, 
and one located at the center. The dissymmetry  between  the
left- and the right-hand sides 
depends on the scaling  parameter $\mu$. The other tunable parameter is the
ratio $R/d$, between the center disk's radius $R$ and the cell's horizontal
width $d$. The radii of the outer disks are fixed to $R/\sqrt{\mu}$ (to the
left) and  $R\sqrt{\mu}$ (to the right). The vertical separations
between the outer disks is taken to be $d\sqrt{3/\mu}$ to the left and
$d\sqrt{3\mu}$ to the right, so that the usual Lorentz channel is retrieved
for $\mu=1$. We note that restrictions must be imposed on the range of
permitted values of the parameters for the self-similar billiard to share
the hyperbolicity of the Lorentz channel. The details can be found in
\cite{BG06}. We assume these conditions to be met in the sequel.

The whole chain is constructed by adding a cell to the right of 
the reference cell, with index $i=1$, identical in shape but with all its
lengths  multiplied by $\mu$, and another one to the left, with index
$i=-1$, with all the lengths divided by $\mu$. We repeat this construction
in such a way that in the $i$th cell, $i\in\mathbb{Z}$, all the lengths are
multiplied by $\mu^i$. The resulting billiard chain is so constructed that
the mirror symmetry with respect to the transformation $\mu\to1/\mu$
remains.


Consider a particle moving inside the billiard with velocity
$\vec{v}$. Figure \ref{fig.channel} shows such a trajectory. 
As the particle moves from one cell to a neighboring one, the length scales
change by a factor $\mu$. Equivalently we can rescale the velocity by a
factor $1/\mu$ and keep the length  scales  unchanged. For both
transformations the characteristic time between successive collisions with
the walls changes by a factor $\mu$. 

We can therefore analyze the dynamics on the extended self-similar billiard
in terms of the dynamics in a periodic cell by applying proper boundary
conditions and rescaling the vertical coordinate and velocity, as the
particle exits either 
\begin{eqnarray}
\mathrm{to\:the\:right~:}&&
\left\{
\begin{array}{lcl@{\quad}c@{\quad}l}
\vec{r} &=& (d,h)&\to&(0,h/\mu),\\
\vec{v} &=& (v_x,v_y)&\to&(v_x/\mu,v_y/\mu),
\end{array}
\right.
\label{pbright}\\
\mathrm{or\:to\:the\:left~:}&&
\left\{
\begin{array}{lcl@{\quad}c@{\quad}l}
\vec{r} &=& (0,h)&\to&(d,h\mu), \\
\vec{v}&=&(v_x,v_y)&\to& (v_x \mu,v_y\mu).
\end{array}
\right.
\label{pbleft}
\end{eqnarray}
Here $h$ is the $y$ coordinate of the trajectory as it crosses the cell's
left or right boundaries. An example is depicted in
Fig.~\ref{fig.unitcell}(b). These conditions are analogous to periodic 
boundary conditions for the usual Lorentz channel billiard, but for the
self-similar structure parametrized by $\mu$. 

Billiards are instances of so-called special flows \cite{CFS82}, which is
to say their time evolutions can be specified by the mapping of the
coordinates between successive collision events, together with the time
interval between them. This kind of representation is particularly useful
to the evolution of self-similar billiards, where the absence of an overall
time scale in the extended system poses difficulties in the definition of
proper time averages. Indeed point particles typically move towards the
direction of increasing cell sizes, thereby decreasing their collision
frequency. Such ambiguities become irrelevant when the evolution 
is considered from one collision event to the next.

The particle's coordinates from one collision with the walls to the next
are specified by the Birkhoff map \cite{Gas98} $\xi_n \equiv (s, v,
\varpi)_n \mapsto \xi_{n+1} \equiv \phi(\xi_n)$. Here the variable $s$
represents the arc-length  along the unit cell boundary (including the open
sides), $v$ is the modulus of the particle's velocity and $\varpi$ is the
sine of the angle between the outgoing velocity and the normal to the
cell's boundary. This map, together with Eqs.~(\ref{pbright}-\ref{pbleft}),
provide the map for the specification of a trajectory.

In order to keep track of the correspondence between the dynamics on the
periodic cell and that on the extended lattice, we introduce a new variable
$I_n$, which takes integer values and labels the cell where the particle is
located after $n$ iterations of the map. This variable changes according to
$I_{n+1} = I_n + a(\xi_n)$, where we introduced the jump function,
$a(\xi_n)$, which takes the values $a(\xi_n)=1$ (resp. $-1$) if
$\phi(\xi_n)$ has the spatial coordinate $s_{n+1}$ on the right
(resp. left) open side of the cell, otherwise $a(\xi_n)=0$.   

The time it takes a trajectory at a (phase space) point $\xi$ on the
boundary to intersect again with the boundary of the billiard depends on
the speed $v$ as  
\begin{equation}
T(\xi)=\frac{L(\xi)}{v}
\label{Time}
\end{equation}
where $L(\xi)$ is the length of the trajectory between intersections at
$\xi$ and $\phi(\xi)$ of the trajectory with the boundary of the unit cell.
A complete characterization of the continuous time flow is thus given by
the variables $(\xi, \tau)$ in the unit cell, with $0<\tau<T(\xi)$, a new
variable that restores the position between collisions. The equivalence
between the continuous time flow on the extended lattice and the Birkhoff
map with the extra variable $\tau$ is further discussed in \cite{BG06}.


A remarkable property of the Birkhoff map defined above is that it does not
preserve phase-space volumes. Indeed, as the particle exits to the
right and re-enters to the left or vice-versa, both $s$ and $v$ coordinates
are rescaled by $\mu^{-1}$ (resp. $\mu$). Accordingly, probability measures
evolve under iteration of the density evolution operator associated to the
Birkhoff map towards a unique invariant measure ({\em i.~e.} a probability
measure invariant under this operator) with fractal properties,
characteristic of a non-equilibrium stationary state. This measure has
three Lyapunov exponents, two negative and one positive, $\lambda_1 > 0
\ge \lambda_2 > \lambda_3$, and such that the phase space contraction rate
$\sigma \equiv - (\lambda_1 + \lambda_2 + \lambda_3) > 0$.

The second Lyapunov exponent, $\lambda_2$, is specific to the self-similar
geometry of the billiard~; it is due to the contraction of the $v$
coordinate as the particle moves around the cell~:  
\begin{equation}
\lambda_2 = -\lim_{n\to\infty}\frac{\Delta I_n}{n} \log \mu \le 0,
\label{lambda2}
\end{equation}
where $\Delta I_n = I_n - I_0$ is the lattice displacement vector after $n$
iterations, or winding number. If $\mu>1$, the particle moves
preferentially to the right, 
corresponding to $\Delta I_n/n>0$ and the other way around if $\mu<1$. 
The symmetric case $\mu=1$ is the usual Lorentz channel with only two
non-trivial Lyapuonv exponents, {\em i.~e.} $\lambda_2=0$. 
We therefore see that the stationary value of $v$ under the Birkhoff map is
trivially $v_n \to v_\infty = 0$, which is to say particles ever move
towards the larger lattice scales, where times between collisions keep
increasing and collisions become seldom.  

$\lambda_2$ accounts for one half of the total phase space contraction
rate. The contraction of the vertical coordinate in
Eqs.~(\ref{pbright}-\ref{pbleft}) accounts for the other half,  
due to $\lambda_1$ and $\lambda_3$, 
\begin{equation}
\lambda_1+\lambda_3 = \lambda_2 < 0.
\label{lambda13}
\end{equation}
These two Lyapunov exponents are related to the components $s$ and
$\varpi$ of $\xi$. 

The invariant measure therefore has a product structure,
$\mathrm{d}^3m(\xi) = \mathrm{d}^2m_1(s, \varpi)\mathrm{d}m_2(v)$. When
$\mu\ne1$, we have $\mathrm{d}m_2(v) = \delta(v)dv$ on the one hand, where
$\delta$ is the Dirac delta function, and, on the other hand, since
$\lambda_1>0$  and $\lambda_3<0$, $m_1$ is a fractal , whose Lyapunov
dimension, here defined by $1 + \lambda_1/|\lambda_3|$, is somewhere
between 1 and 2. 



The entropy production per collision can be defined dynamically as the
phase space contraction rate \cite{R96}, $\sigma = - (\lambda_1 + \lambda_2
+ \lambda_3) = -2\lambda_2$, which can be computed in terms of the winding
number, as in Eq.~(\ref{lambda2}). 

The winding number can furthermore easily be connected 
to the average drift of particles on the extended system. As we showed in
\cite{BG06}, the average of the spatial coordinate $X$ on the extended
system evolves linearly in time, with a slope $V$ which defines the drift
velocity, $\langle X \rangle_t = V t$. The winding number and drift
velocity are equal up to dimensional factors~:
\begin{equation}
\lim_{n\to\infty} \frac{\Delta I_n}{n} = \frac{T_\mu}{d} V,
\label{windingdrift}
\end{equation}
where $d$ is the reference cell's width, and $T_\mu$ the average time
between collisions measured in that same cell.
The phase space contraction rate can therefore
be expressed directly in terms of the average drift velocity~:
$
\sigma = \frac{2 T_\mu V}{d} \log \mu
$.
This expression can be transposed to a phase space contraction rate per
unit time in the limit $\mu\to1$. In this limit, $V$ and $\log\mu$ both
scale like $\mu-1$, so that the lowest order contribution to $\sigma$ is
$\mathcal{O}(\mu-1)^2$, with $T_{\mu=1}$. Divinding $\sigma$ by the latter
defines the (dynamical) entropy production per unit time,
\begin{equation}
\frac{\sigma}{T_{\mu=1}} = \frac{2 V}{d} (\mu - 1) + \mathcal{O}(\mu-1)^3.
\label{psceptime}
\end{equation}

As we showed in \cite{BG06}, the phenomenology of self-similar systems such
as this one is characterized on the one hand by a constant drift from the
small to the large scales and, on the other, by a mean square displacement
which undergoes a transition from diffusive to ballistic behavior at a
crossover time which scales like $\mu^{3/2}/(\mu-1)$. Therefore, as
$\mu\to1$, self-similar systems can be considered as diffusive systems with
constant drift, such as modeled by a Smoluchowski type equation for the
probability density $W$, $\partial_t W = \mathcal{D}\partial^2_x W - V
\partial_x W$. In the stationary state, the phenomenological entropy
production rate is given \cite{VTB97} by the drift velocity $V$ squared,
a term quadratic in $\mu - 1$, divided by the diffusion coefficient
$\mathcal{D}$, here computed at $\mu = 1$. 

\begin{figure}[htb]
\centering
\includegraphics[width=.45\textwidth]{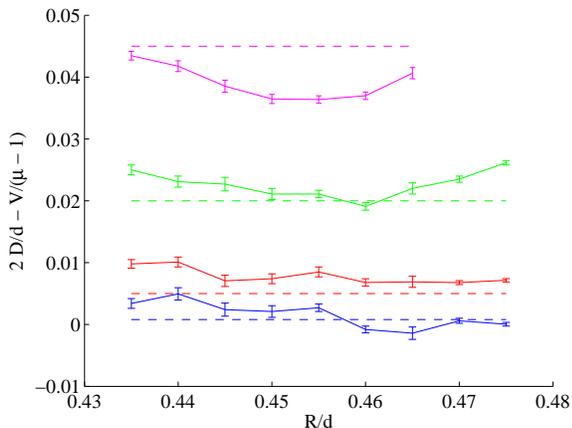}
\caption{(Color online) The difference $2\mathcal{D}/d - V/(\mu-1)$
  vs. $R/d$ is $\mathcal{O}(\mu-1)^2$, as expected from
  Eq.~(\ref{vd}). Here shown are the values $\mu =
  1.02,\:1.05,\:1.10,\:1.15$ from bottom  to top. The straight  lines show
  $2(\mu-1)^2$.  Here $V$ is 
  computed as the slope of $\langle X \rangle_t$ vs. $t$, where the average
  is computed out of many random initial conditions.}  
\label{fig.ep}
\end{figure}

Equating Eq. (\ref{psceptime}) with the expression of the thermodynamic
entropy production rate at $\mathcal{O}(\mu-1)^2$, we infer the following
relationship between the velocity drift and diffusion coefficient~:
\begin{equation}
\lim_{\mu\to1}\frac{V}{\mu-1} = \frac{2\mathcal{D}}{d},
\label{vd}
\end{equation}
which can be viewed as a definition of the mobility, the external field
being given by $2(\mu-1)$, and therefore establishes the equivalent of a
fluctuation-dissipation theorem. This relationship is confirmed by the
numerical results presented in Fig. \ref{fig.ep}, which were carried out
for different choices of $R/d$ and four different values of $\mu$ close to
$1$. Equation~(\ref{windingdrift}) was also verified in the numerical 
scheme.

To conclude, self-similar billards are simple mechanical models of
conduction in spatially extended systems with volume-preserving dynamics,
whereby a geometric constraint induces a steady current.  As in the
conformal transformation of the GIKLG \cite{BG}, or the multibaker maps with
energy \cite{TG99}, the essential ingredient for this behavior is the rapid
increase of phase-space volumes, whose rate is here given by $\mu$.

These models lend themselves to further exploring the connections between
the chaotic properties of non-equilibrium dynamical systems and
phenomenological thermodynamics. The dynamical properties are conveniently
analyzed by the means of the Birkhoff map on the periodic cell, which
specifies the evolution of phase points from one collision to the next,
here on a three-dimensional space. 

Whereas the dynamics on the extended system preserves phase-space volumes,
the Birkhoff map on the periodic cell contracts phase-space volumes, with an 
invariant measure whose fractal dimension is between 1 and 2. A similar
mechanism of contraction of phase-space volumes at the periodic boundaries
occurs with the W-flow. The phase-space contraction rate can 
furthermore easily be computed in terms of the map's winding number, which
bears a simple connection to the average velocity drift associated to the
macroscopic current. The comparison between phase-space contraction rate and
phenomenological entropy production yields an expression of the velocity
drift near the equilibrium regime.

\begin{acknowledgments}
The authors are grateful to N.~I. Chernov and C. Liverani for helpful
comments. FB acknowledges financial support from Fondecyt
project 1060820 and FONDAP 11980002. TG  is financially supported by the
Fonds National de la Recherche Scientifique. This collaboration was
supported through grant ACT 15 (Anillo en Ciencia y Tecnologia).   
\end{acknowledgments}


\begin{thebibliography}{99}

\bibitem{H86} W. G. Hoover, {\em Time Reversibility, Computer Simulation,
    and Chaos}, (World Scientific, Singapore, 2001).

\bibitem{EM90} D. J. Evans and G. P. Morriss, {\em Statistical Mechanics of
    Non-Equilibrium Fluids} (Academic Press, London, 1990).

\bibitem{Gas98} P. Gaspard, {\em Chaos, Scattering and Statistical
    Mechanics} (Cambridge University Press, Cambridge, 1998). 

\bibitem{Dor99} J. R. Dorfman, {\em An Introduction to Chaos in
    Nonequilibrium Statistical Mechanics} (Cambridge University Press,
  Cambridge, 1999). 

\bibitem{CELS93} N. I. Chernov, G. L. Eyink, J. L. Lebowitz, and
  Ya. G. Sinai, 
  Phys. Rev. Lett. {\bf 70} 2209 (1993); 
  Comm. Math. Phys. {\bf 154} 569 (1993).  

\bibitem{R96} D. Ruelle, 
  J. Stat. Phys.{\bf 85} 1 (1996); 
  Proc. Nat. Acad. Sci. {\bf 100} 30054 (2003); 
  Physics Today {\bf 57}, 5, 48 (2004).  


\bibitem{DM98} C.~P. Dettmann and G.~P. Morriss, Phys. Rev E {\bf 54} 2495
  (1996); G.~P. Morriss and C.~P. Dettmann, Chaos {\bf 8} 321 (1998).

\bibitem{W00} M.~P. Wojtkowski, J. Math. Pures Appl. {\bf 79} 953 (2000).

\bibitem{BG} F. Barra and T. Gilbert, J. Stat. Mech. L01003 (2007).

\bibitem{BG06} F. Barra, T. Gilbert, and M. Romo, 
  Phys. Rev. E {\bf 73} 026211 (2006).

\bibitem{BR01} G. Benettin and L. Rondoni, MPEJ {\bf 7} 3 (2001).

\bibitem{TG99} S. Tasaki and P. Gaspard,
Theor. Chem. Acc. {\bf 102} 385 (1999); 
J. Stat. Phys. {\bf 101} 125 (2000).

\bibitem{CFS82} I. P. Cornfeld, S. V. Fomin and Ya. G. Sinai,
  {\em Ergodic Theory} (Springer-Verlag, Berlin, 1982).

\bibitem{VTB97} J. Vollmer, T. T\'el, and W. Breymann,
  Phys. Rev. Lett. {\bf 79} 2759 (1997); 
  Phys. Rev. E {\bf 58} 1672 (1998);
  Chaos {\bf 8}, 396 (1998).

\bibitem{MDR97} G. P. Morriss, C. P. Dettmann, and L. Rondoni,
  Physica A {\bf 240} 84 (1997).




\end{thebibliography}
\end{document}